\documentstyle[12pt,aasms4]{article}

\received{May 2000}
\accepted{October 2000}

\slugcomment{to appear in the Astrophysical Journal Letter}

\lefthead{Tashiro et al.}
\righthead{Field and Particle Energy Distributions in the Fornax A West Lobe}

\begin{document}

\title{X-Ray Measurements of the Field and Particle Energy Distributions\\
	in the West Lobe of the Radio Galaxy NGC~1316 (Fornax~A)}

\author{M. Tashiro\altaffilmark{1}, 
        K. Makishima\altaffilmark{2}, 
	N. Iyomoto\altaffilmark{3}, 
	N. Isobe\altaffilmark{2}, 
 and    H. Kaneda\altaffilmark{3}}
\altaffiltext{1}{Department of Physics, Saitama University, Shimo-Okubo, 
		Urawa, 338-8570, Japan. MT: tashiro@phy.saitama-u.ac.jp}
\altaffiltext{2}{Department of Physics, University of Tokyo, Hongo, 
		Bunkyo, 113-0033, Japan.}
\altaffiltext{3}{Institute of Space and Astronautical Science,
		Yoshinodai, Sagamihara, 229-8510, Japan.}

\begin{abstract}
A follow-up X-ray study was made of the west lobe of the radio galaxy
Fornax~A, (NGC~1316) based on new {\it ASCA} observations made in 1997
for 98 ks, and incorporating the previous observation in 1994 for 39
ks.  The 0.7--10 keV spectrum of the emission can be described by a
power-law of energy index $0.74\pm0.10$, which agrees with the
synchrotron radio index of $0.9\pm0.2$.  Therefore, the X-rays are
reconfirmed to arise via inverse-Compton scattering of the cosmic
microwave photons, as Kaneda et al. (1995) and Feigelson et al. (1995)
concluded.  The surface brightness of the inverse-Compton X-rays
exhibits a relatively flat distribution over the west lobe, indicative
of an approximately spherical emissivity distribution with a radius of
$\sim11'$ (75 kpc).  In contrast, the 1.4 GHz radio image by Ekers et
al. (1983) exhibits a rim-brightened surface brightness, consistent
with a shell-like emissivity distribution whose inner and outer
boundaries are $4'$ and $11'$, respectively.  These morphological
differences between radio and X-rays suggest that the relativistic
electrons are distributed homogeneously over the lobe volume, whereas
the magnetic field is amplified toward the lobe rim region.
\end{abstract}


\section{Introduction}
The jet terminal lobes of radio galaxies are gigantic inter-galactic
structures consisting of magnetic fields and relativistic particles,
both of which are supposed to be supplied by the active galactic
nuclei (AGNs) through the jets.  Nevertheless, we do not yet know the
relative importance of the magnetic fields and the relativistic
particles in the mechanism of jet formation.  To answer the issue, we
need to measure spatial distributions of the field and particle energy
densities along various locations of the AGN-jet-lobe system.

The relativistic electrons in the radio lobes interact with the
magnetic fields and soft photons to produce synchrotron radiation and
inverse-Compton (IC) emission, respectively.  Feigelson et al. (1995,
hereafter FEA95) and Kaneda et al. (1995, KEA95) discovered the IC
X-ray emission with {\it ROSAT} (Tr\"{u}mper 1983) and {\it ASCA}
(Tanaka et al. 1994), respectively, from the lobes of the radio galaxy
Fornax~A (NGC~1316, redshift $z = 0.00587$; Longhetti et al. 1998),
followed by authors reporting the lobe IC X-rays from Centaurus~B
(PKS~1343$-$601; Tashiro et al. 1998), 3C~219 (Brunetti et al. 1999)
and NGC~612 (Tashiro et al. 2000).  These authors estimated the energy
density of soft photons, and succeeded in determining the magnetic
field intensity and the energy densities of electrons in the lobes.
Among these pioneering results, Tashiro et al. (1998) and Brunetti et
al.  (1999), respectively, showed that the lobes of Centaurus~B and
3C~219 exhibit clear dominance of particle energy, in their inner
regions. Notably, Tashiro et al. (1998) found relative enhancements of
the magnetic fields towards the periphery in the Centaurus~B lobes.

Our next task is to resolve spatial distributions of the IC X-ray
emission to trace the dynamics of the jet-lobe system.  The lobes of
Centaurus~B, however, are not in fact the best target for this
investigation with {\it ASCA}, because of the bright nucleus and the
relatively small angular separation of its lobes.  In this paper, we
present results from {\it ASCA} follow up observations of the west
lobe of Fornax~A.  This prototypical IC source is ideal for our
purpose, because of its nearly dormant nucleus (Iyomoto et al. 1998),
the large ($\sim 10'$ in radius) angular size of its lobes, and the
high integrated IC flux.

\section{Observation} 
In the first {\it ASCA} observation of Fornax~A carried out on January
11, 1994, KEA95 placed the nucleus at the center of the field of view
(fov).  This was less efficient to observe the lobes because of the
vignetting effect of the X-Ray Telescope (XRT: Serlemitsos et
al. 1993), and, furthermore, some portion of the lobes fell outside
the fov. Aiming at the detail study of the lobe, we conducted follow
up observations of Fornax~A by placing the west lobe, which is less
contaminated by point sources (FEA95; KEA95; see also Kim et al. 1999)
than the other, in the center of the Gas Imaging Spectrometer (GIS:
Ohashi et al. 1996; Makishima et al. 1996) fov.  We concentrate on the
GIS data, since a fair amount of the lobe emission remained outside
the $22'\times11'$ fov of the Solid-state Imaging Spectrometer (SIS:
Burke et al. 1991; Yamashita et al. 1997) at the {\it 2-CCD} mode.
The observations were carried out on August 17--18 and December 26 --
27, 1997.  The good time exposure after standard data screening is 35
ks and 62 ks, for the August and December observations, respectively.
Including the observation in 1994, the total exposure time amounts to
137 ks, which is more than four times as long as that obtained by
KEA95.

\section{Retults}
\subsection{Imaging analysis}
In Figure~1, we show 0.7 -- 10 keV GIS images obtained in the two
observations in 1997.  We smoothed the raw images with a 2-dimensional
gaussian function of $\sigma = 0.'5$, but left cosmic and intrinsic
backgrounds in these images (Fig.1).  Both two images show the diffuse
IC emission over the west lobe region, as first revealed by FEA95 and
KEA95.  The brightest and the second brightest discrete sources are,
respectively, the host galaxy NGC~1316 and the SB galaxy NGC~1310.
Besides a new transient source appeared on the second occasion to the
southwest of the lobe, we see no significant variation of the sources
in the fov.  We therefore co-add the GIS (GIS2+GIS3) data from the
present observations with those from the first one made in 1994 in
order to study the IC X-ray emission with the best signal-to-noise
ratio.  For that purpose, we subtracted the intrinsic non-X-ray
background (NXB) utilizing night earth exposures close in time to each
observation, corrected the image (containing the cosmic X-ray
background, CXB, and the sources) for the exposure and vignetting
effect, and smoothed it with a 2-dimensional gaussian function of
$\sigma =0.'4$.

The synthesized X-ray map is shown in Figure~2 with gray scales, where
the 1.4 GHz radio contours by Ekers et al. (1983) are overlaid.  We
detected four additional point-like source candidates (labeled 1 -- 4)
with a threshold of 4 $\sigma$ deviation above the average flux over
the lobe region.  Apart from these small-scale features, we see the
diffuse IC emission exhibiting a good coincidence with the radio lobe.
We analyzed the {\it ROSAT}-PSPC (Pfeffermann et al. 1986) archival
data of these regions, and found that all the sources but the
transient source were identified with the {\it ROSAT} data as point
sources. The extrapolated 2 -- 10 keV flux is higher than
$2\times10^{-17}$W~m$^{-2}$ from their spectrum fitting results, and
the derived energy indices range 1.5 -- 3.2.  These soft X-ray spectra
imply that the discrete sources do not originate from the possible
local enhancements of the diffuse IC emission.  We examined {\it ASCA}
data for NGC~1310 and source 1 -- 4, and confirmed their intensities
stayed constant within errors among these observations with {\it ASCA}
and {\it ROSAT}.  Although the sources 1 and 2 are included in the
region employed by KEA95 for their spectrum analysis, the sum flux of
these sources is not more than 1/10 of the flux reported by KEA95.
The sources little affect their conclusion, assuming them to be
stable.

\subsection{Spectrum of the diffuse emission}
We accumulate the GIS events over a circular region (dashed circle in
Fig.~2) around the X-ray brightness peak (marked with a cross in
Fig.~2). We limited the area within $7'.25$ to avoid source
contamination from the host galaxy NGC~1310 and source 2, although the
source 1 is inevitably contained in the integration region.  Since
there is essentially no source-free region in the on-source fov, we
estimated the background (NXB + CXB) utilizing archival blank-sky
observations accumulated for 1653 ks.  Furthermore, we decomposed the
background spectrum into the NXB and CXB components referring to the
night earth data, and rescaled the NXB normalization to take into
account the secular change in the NXB, by comparing the night earth
spectrum between the epoch of the on-source observation (in 1994 or
1997) and that of the blank-sky archival data (mostly obtained in
1993).  We thus generated an appropriate background spectrum (CXB plus
rescaled NXB) for each on-source observation, and subtracted it from
the raw spectrum of each GIS instrument.  Then we summed the GIS
spectra from the observations into a single one.

In Figure~3, we show the obtained background subtracted GIS2 + GIS3
spectrum of the west lobe region.  It is relatively hard and
significantly detected up to $\sim 8$ keV, implying a considerable
improvement over results in KEA95.  We fitted the spectrum with a
model consisting of three components; (1) a power-law with free energy
index and free normalization, representing the IC emission; (2) a
thin-thermal emission (Raymond \& Smith 1977) with free temperature
and free normalization, but with the metallicity fixed at 0.4 solar
abundance, which represents the soft thermal emission surrounding the
radio galaxy as detected by KEA95; and (3) the contaminating source 1.
We analyzed the {\it ROSAT}-PSPC spectrum of source 1, and found that
it can be described with a power-law of energy index $3.2\pm0.7$ and
flux density $0.008 \pm 0.001$ $\mu$Jy [$ = (8\pm1) \times
10^{-35}$W~m$^{-2}$Hz$^{-1}$] at 1~keV, absorbed by a column density
of $(8.0\pm0.2)\times 10^{24}$ H~atoms~m$^{-2}$ with $\chi^2 /{\rm
d.o.f.} = 12.1 / 17$.  In the fit to the GIS data, we constrained the
component (3) to take these best-fit {\it ROSAT} parameters.  We
calculated the {\it ASCA} response function for the three observations
individually, and then took their weighted average.  This three
component model has successfully described the spectrum with $\chi^2
/$d.o.f. $= 94.7 / 103$.  The best-fit parameters are shown in Table
1.  The 2 -- 10 keV flux obtained from the lobe region is $4.8 \times
10^{-16}$W~m$^{-2}$, which is consistent with that evaluated by KEA95. 
The derived energy index of the IC component ($0.74\pm0.26$) agrees
with that obtained by KEA95 ($1.4\pm0.7$), and the accuracy is much
improved.

A similar good fit was obtained by replacing the power-law component
(1) with a second thin thermal plasma emission model, of temperature
$kT = 8.4^{+\infty}_{-3.2}$ keV, emission measure of $\int n_{\rm
e}n_{\rm H} dV = (6.6\pm1.2) \times 10^{56}$m$^{-3}$, and abundance
fixed at 0.4 solar.

\subsection{X-ray brightness distribution}
We saw in Figure~2 that the IC X-ray emission is relatively
concentrated on the center of the lobe.  To quantify the surface
brightness distributions, we masked the contaminant discrete sources
(\S~3.1) and made a 1.0 -- 10 keV radial brightness profile around the
X-ray brightness peak (the cross in Fig.~2).  The X-ray surface
brightness distribution is so flat within a few arcmin that the radial
profile is little sensitive to the position of the center (Fig.~2).
To compare with the X-ray radial profile around the same center, we
made a radial brightness profile in radio utilizing the 1.4 GHz image
by Ekers et al. (1983), and show them in Figure~4.  Thus, the two
profiles resemble each other, but differ in two points; the
rim-brightened feature seen in the radio profile is absent in X-rays,
and the X-ray distribution appears to have a larger radius than the
radio distribution.

To examine possible instrument artifacts, we simulated an expected
X-ray image with the XRT and GIS response simulator to fit the
observed radial brightness profile.  The position resolution and
vignetting effect of were calculated based on calibration data from
Cygnus X-1 (Ikebe 1995; Takahashi et al. 1995).  We assume isotropic
distribution of the emissivity through the simulation.  This analysis
revealed that a homogeneously filled spherical emissivity model with
the outer boundary of $(12\pm1)'$ reproduces the observed X-ray radial
profile very well (the solid line in the Fig.~5).

On the other hand, a shell-shaped emissivity model reproduced the rim
brightened radio profile very well, as represented with a dashed line
in Figure~4.  The outer boundary of the shell is constrained to be
$(11\pm0.5)'$, while the inner boundary $(4\pm0.5)'$.  The
shell-shaped model, however, could not describe the obtained X-ray
profile with a finite inner boundary.  Therefore, unlike the case of
X-ray emissivity, the radio profile requires a reduced emissivity at
the lobe center, although we see no significant discrepancy between
the derived outer boundaries of the X-ray and radio profiles. Since
the radio data provided better accuracy, we regard the value of $\sim
11'$ derived from radio data as the common outer boundary of radio and
X-rays in discussion below.

\section{Discussion}
We measured the diffuse X-rays of the Fornax~A west lobe with an
improved accuracy employing the {\it ASCA} follow up observations.
The diffuse emission spatially coincides with the radio synchrotron
lobe on a large scale.  The derived 0.7--10 keV spectrum of the lobe
X-rays is well described with a power-law model, whose energy index of
$0.74\pm0.26$ agrees with the synchrotron radio emission energy index,
$0.9\pm0.2$, derived from published radio fluxes observed at 408 MHz,
1.4 GHz, and 2.7 GHz (Cameron 1971; Ekers et al. 1983; Shimmins 1971,
respectively).  Although the obtained spectrum formally accepts a thin
thermal plasma model, the high temperature ($kT =
8.4^{+\infty}_{-3.2}$ keV) will allow neither gravitational
confinement nor cooling flow to explain the emission.  The derived
emission measure requires a thermal pressure of $\sim 3\times10^{-13}$
J~m$^{-3}$.  If we assume a magnetic field of $\sim 1$ nT to confine
the possible thermal plasma, that field will reduce the synchrotron
cooling time down to $\sim 10^6$ yr (Iyomoto et al. 1998).  This is
too short for the relativistic electrons to diffuse in the whole
lobes, considering the estimated growth speeds of lobes (Scheuer
1995).  Therefore, we reconfirm, with a higher accuracy, that the
diffuse X-ray emission is produced via the IC process in which the
cosmic microwave background (CMB) photons are boosted by the
synchrotron electrons, as FEA95 and KEA95 concluded.

With the follow up observations pointing on the west lobe, we examined
the brightness distribution of the IC X-ray emission, and revealed
that it indicates a center-filled emissivity distribution in contrast
to the rim-brightening radio shape.  This indicates that the
relativistic electrons homogeneously in the sphere of radius $\sim 75
h_{75}^{-1}$ kpc $(= 11')$, where $h_{75}$ is the Hubble constant
normalized to $75 \:\:{\rm km~s}^{-1} {\rm Mpc}^{-1}$.  We therefore
regard the rim brightened radio emission as a direct indicator of the
magnetic pressure distribution, represented with a shell of inner and
outer boundaries of $\sim 27 h_{75}^{-1}$kpc $(=4')$ and $\sim 75
h_{75}^{-1}$kpc $(=11')$, respectively.  On the basis of the spectral
and spatial analysis, we evaluate the energy densities of magnetic
field ($u_{\rm B}$) and electrons ($u_{\rm e}$) according to Harris
and Grindlay (1979).  Since $u_{\rm B}$ is determined by the ratio of
the synchrotron radio flux to the IC X-ray flux emitted from the same
volume, $u_{\rm B}$ is in inverse proportion to the volume ratio of
the shell to the electron filled sphere, which is $\sim 0.95$.  We
estimate the 1.4 GHz flux falling inside the X-ray integration region
as 50 Jy ($=5\times10^{-25}$W~s$^{-1}$m$^{-2}$), based on the radio
map by Ekers et al. (1983).  We then calculate $u_{\rm B}$ after
Harris and Grindlay (1979), considering the filling factor ($f$) to
the shell region, and assuming the magnetic fields to have random
directions to the line of sight. The result is $u_{\rm B} = (4.6 \pm
0.4) f^{-1} \times10^{-14}$ J~m$^{-3}$.

The derived $u_{\rm B}$ corresponds to the magnetic field strength of
$(0.34\pm0.02)$ nT at $f = 1$.  Note here we adopted an IC scattering
electron density, at Lorentz factor $\gamma \sim 1000$, estimated
through extrapolation from the observed synchrotron spectrum at
$\gamma \sim 5000$ -- 13000.  Utilizing equation (10) of Harris and
Grindlay (1979), we also derive $u_{\rm e}$ in the sphere.  We assume
that the electron Lorentz factor ranges $\gamma = 10^{3-5}$ in
calculation below, as KEA95 adopted.  Thus we obtained $u_{\rm e}$
from the estimated IC X-ray flux (Table 1) as 
$u_{\rm e} = (2.0 \pm 0.5) (r_{\rm out}/11')^{-3} h_{75} 
					\times 10^{-14}$ J~m$^{-3}$.
Although the derived $u_{\rm B}$ is nominally larger than the $u_{\rm
e}$ by a factor of 2.3 in the shell, we cannot immediately reject
electon-magnetic field equipartition ($u_{\rm B} = u_{\rm e}$) there,
without determining the electron energy spectrum below 
$\gamma \sim 5000$.  If the spectrum turns down just below 
$\gamma = 5000$, the estimated $u_{\rm B} / u_{\rm e} \gtrsim 8.3$.  
On the contrary, assuming that the spectrum extends down to 
$\gamma \sim 350$ ($\sim 2$ MHz in synchrotron emission), the integrated 
$u_{\rm e}$ becomes equivalent to $u_{\rm B}$ in the shell region.

The results obtained here suggest a picture that the lobe interior is
dominated by the particle pressure, whereas the magnetic field
pressure becomes significant in the shell region defined by the lobe
boundary.  Interestingly, we see a similar situation in the lobes of
Centaurus~B (PKS~1343$-$601), as Tashiro et al. (1998) showed. The
similarity is thought to reflect energetics and evolutions of radio
lobes (e.g. Blundell \& Rawlings 2000) and the inter-galactic
environments of particle and magnetic fields.  Further investigation
with recent advanced X-ray observatories are expected.

\clearpage

\begin{table*}
\begin{center}
\tablenum{1}
\caption{Results of the model fit to the 0.7 -- 10 keV GIS spectrum from the lobe region
\tablenotemark{a}}

 \begin{tabular}{lcccc} 
\hline \hline 
component
	& $N_{\rm H}$\tablenotemark{b}  
		& energy index 
			&  $kT$ [keV]  
				&$F_{\rm 1keV}$\tablenotemark{c}\\
\hline 
source 1& (8.0)\tablenotemark{d} & (3.2)\tablenotemark{d}&  ---  
				& (0.008)\tablenotemark{d}\\
IC emission & (2.06)\tablenotemark{e} & $0.74\pm0.26$ &  --- & $0.10\pm0.01$ \\
ambient thermal &(2.06)\tablenotemark{e}& --- & $0.85\pm0.15$& $0.11\pm0.05$\\
\hline 
\end{tabular} 
\tablenotetext{a}{The fit chi-square is 94.7 for 103 degree of freedom. 
	All the errors refer to single-parameter $90 \%$ confidence limits.}
\tablenotetext{b}{Photoelectric absorption column density in the unit of 
	$10^{24}$H atoms~m$^{-2}$.}
\tablenotetext{c}{Flux density at 1 keV in the unit of 
	1 $\mu$Jy = $1\times 10^{-32}$W~m$^{-2}$Hz$^{-1}$.}
\tablenotetext{d}{Fixed at the best-fit values of the {\it ROSAT}-PSPC data.}
\tablenotetext{e}{Fixed at the Galactic line-of-sight value.}
\end{center}
\end{table*}

\clearpage

\begin{figure}
\epsscale{1.0}
\plotone{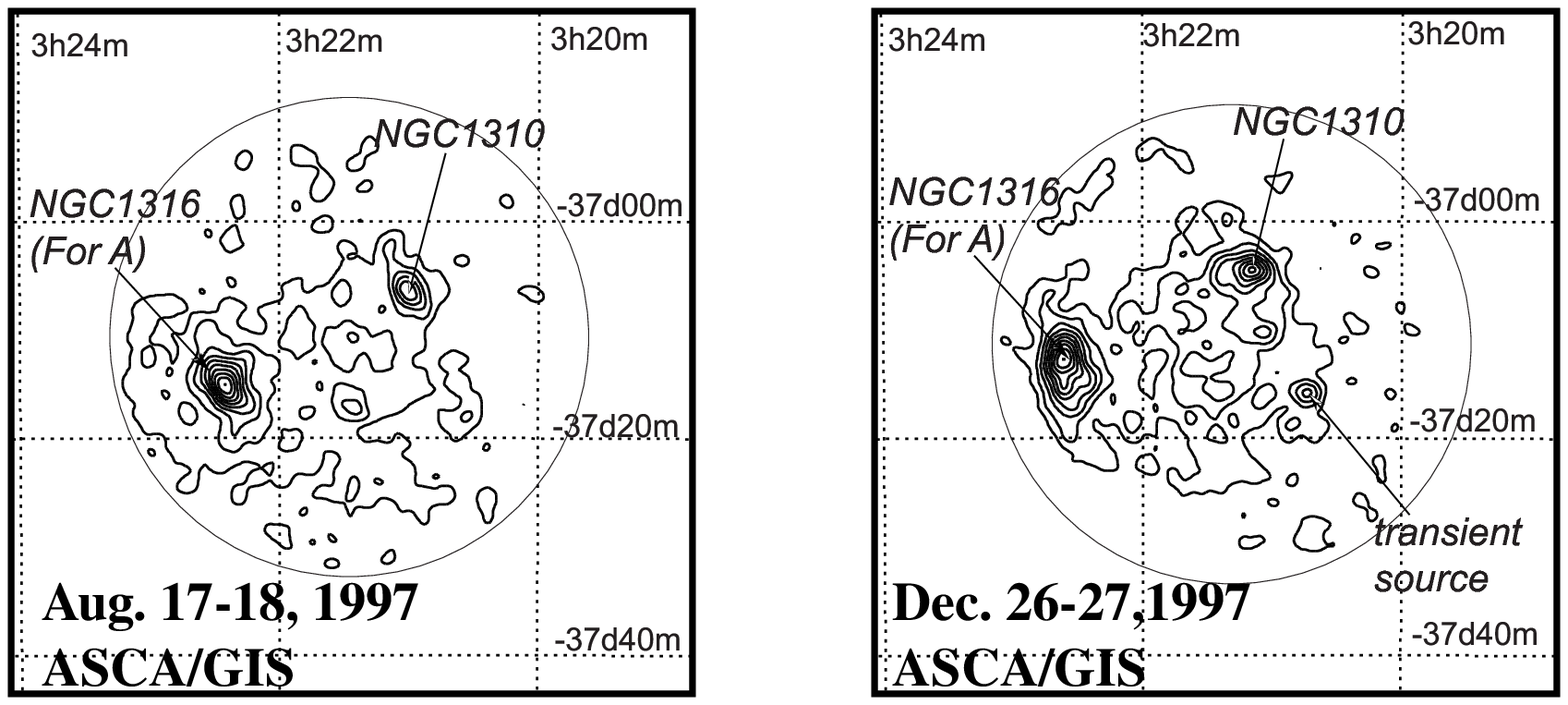} 
\caption{Background-inclusive GIS images obtained on August 15--16
(left panel) and December 26--27 in 1997 (right panel), excluding the
rim region beyond $20'$ from the center of field of view.  Images are
smoothed with a two dimensional gaussian function of $\sigma = 0.'5$,
but not corrected for exposure or vignetting.  Ten linear contour
levels, including backgrounds, range $(1.7$ -- $11)\times 10^{-5}$
counts~s$^{-1}$ and $(1.7$--$8.9)\times 10^{-5}$ counts~s$^{-1}$ for
August and December data, respectively.}
\end{figure}

\begin{figure}
\epsscale{0.7}
\plotone{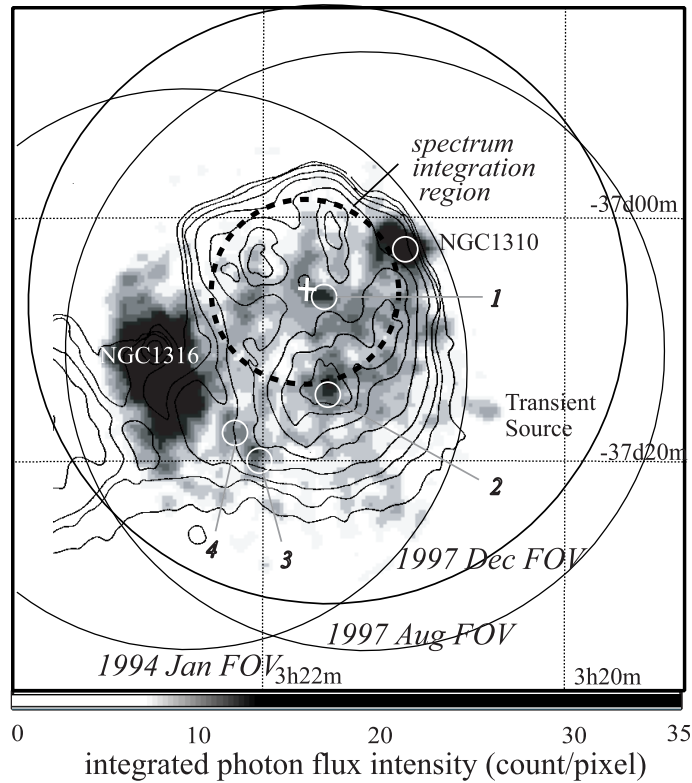} 
\caption{The gray scale shows Fornax~A in a synthesized image (0.7 --
10 keV), obtained by co-adding GIS2 and GIS3 data from the three
pointings, normalized to the exposure and corrected for vignetting
after subtracting the intrinsic background.  The fields of view of
individual pointings (see \S2) are indicated with circles.  The 1.4
GHz radio image from Ekeres et al. (1983) is overlaid with contours.
The dashed-line circle within the west lobe indicates the region used
to accumulate the GIS spectrum (see \S3.2 and Fig.~3).  White circles
represent discrete sources detected in the west lobe with {\it ROSAT}
whose extrapolated 2 -- 10 keV fluxes exceed $2\times10^{-17}$
W~m$^{-2}$ (\S~3.1).}
\end{figure}

\begin{figure}
\epsscale{0.7}
\plotone{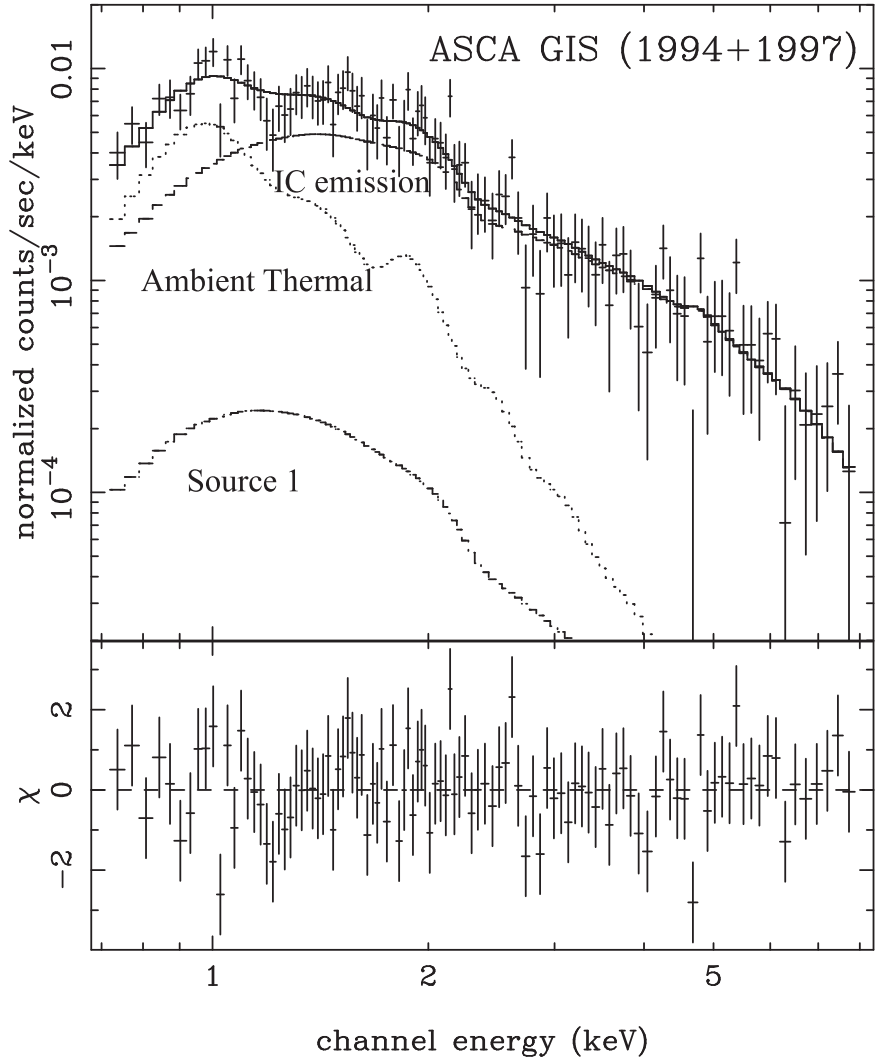} 
\caption{ASCA-GIS spectrum (0.7 -- 10 keV) accumulated from the
integration region shown in Fig.1. Three spectra from the three
pointings were co-added.  The cosmic and intrinsic backgrounds were
subtracted from each spectrum, taking into account the secular change
of the non-X-ray background (see text). The best-fit model and its
constituent components are shown, and the best-fit parameters are
given in Table 1.}
\end{figure}

\begin{figure}
\epsscale{0.7}
\plotone{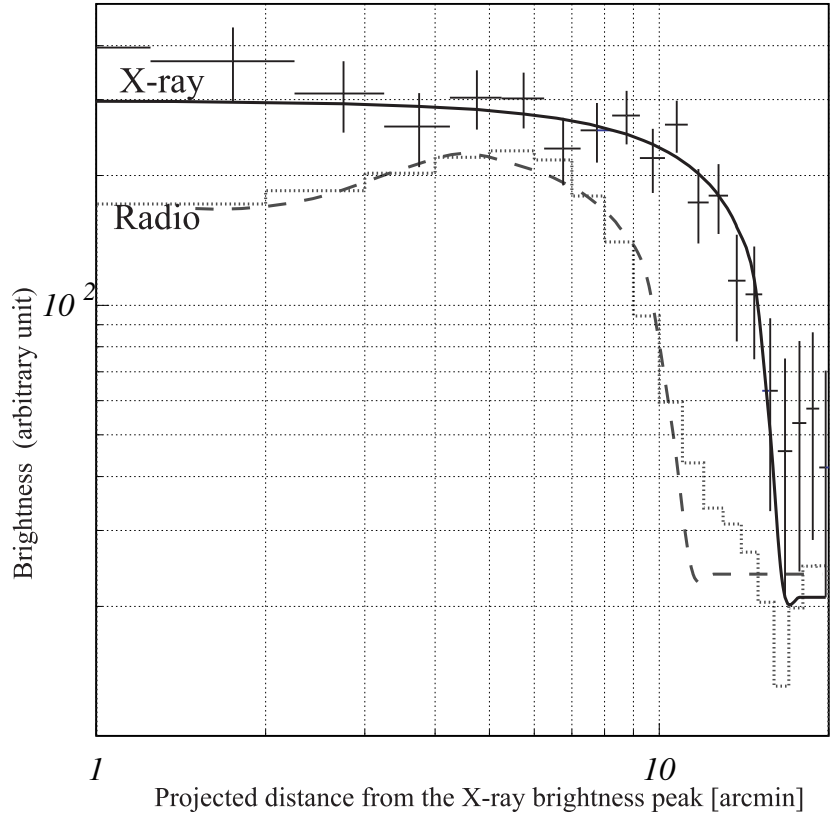}
\caption{Radial X-ray (0.7 -- 10 keV) and radio surface brightness
distributions measured from the west radio lobe of Fornax~A.  The
abscissa is projected distance from the X-ray peak in the unit of
arcmin, while the ordinate is the relative brightness in an arbitrary
unit.  The X-ray data (cross) are obtained from the ASCA observations
in 1994 and 1997 after subtracting background, while the radio
distribution (dotted histogram) is compiled from the map presented by
Ekers et al. (1983).  Fitted model distributions are also indicated.
Solid and dashed lines represent a filled sphere and a shell models,
fitted to the X-ray and radio profiles, respectively (see text).}
\end{figure}

\end{document}